\documentclass[12pt,a4papers,floats]{article}
\usepackage{amsmath,amsfonts}
\makeatletter \@addtoreset{equation}{section}

\usepackage{bm}
\usepackage{dcolumn}
\makeatother
\def\one{{\hbox{ 1\kern-.8mm l}}}
\newcommand{\Dslash}{\not{\hbox{\kern-4pt $D$}}}
\newcommand{\pdslash}{\not{\hbox{\kern-2pt $\partial$}}}
\newcommand{\be}{\begin{equation}}
\newcommand{\bea}{\begin{eqnarray}}
\newcommand{\eea}{\end{eqnarray}}
\newcommand{\ba}{\begin{array}}
\newcommand{\ea}{\end{array}}
\newcommand{\ee}{\end{equation}}

\newcommand{\nuo}{{r}_++{r}_--{1\over\nu}\sqrt{{r}_+{r}_-(\nu^2+3)}}
\begin{document}
\begin{titlepage}
\vspace*{1mm}%
\hfill%
\vbox{
   \halign{#\hfil        \cr
          IPM/P-2010/023 \cr
%          % SUT-P-07-2b   \cr
                     } % end of \halign
     }  % end of \vbox
\vspace*{10mm}%
\begin{center}

{{\Large {\bf Hidden Conformal Symmetry of  Warped }}}\\
\vspace*{5mm}
 {{\Large {\bf
AdS$_3$ Black Holes  }}}

\vspace*{10mm} \vspace*{1mm} {{ Reza Fareghbal }}

 \vspace*{1cm}
{\it  School of physics, Institute for Research in Fundamental Sciences (IPM)\\
P.O. Box 19395-5531, Tehran, Iran \\ }
%
%\vspace*{.4cm}
%
%
%{\it ${}^b$ Department of Physics, Sharif University of Technology \\
%P.O. Box 11365-9161, Tehran, Iran}
%
%
\vspace*{.4cm} {  fareghbal@theory.ipm.ac.ir}

\vspace*{1cm}
%%%\maketitle
\end{center}
\begin{abstract}
We show that for a  certain low frequency limit, the wave equation of a generic  massive  scalar field in the background of the spacelike  warped AdS$_3$ black hole can be written as the Casimir of an $SL(2,R)$ symmetry.
Two sets of $SL(2,R)$ generators are found  which uncover the hidden $SL(2,R)\times SL(2,R)$ symmetry of the solution. This symmetry is only  defined locally and is  spontaneously broken  to $U(1)\times U(1)$  by a periodic identification of the $\phi$ coordinate. By using the generator  of the identification we  read the left and right temperatures $(T_L,T_R)$ of the proposed dual conformal field theory which are in complete agreement  with the $WAdS/CFT$ conjecture. Moreover,   under the above condition of the scalar wave frequency, the absorption cross section of the scalar field is consistent with the  two-point function of the dual CFT.
\end{abstract}
\end{titlepage}
%--------------------------------------------------------------------
%Introduction
\section{Introduction}
It is well-known that adding higher derivative terms to  three dimensional gravity (with cosmological constant) yields to novel theories. An interesting example of such theories is Topologically Massive Gravity (TMG) \cite{Deser:1981wh,Deser:1982vy}. It contains the gravitational Chern-Simons action as a correction to the 3D Einstein gravity with the cosmological constant (see \eqref{TMGL} ). Besides the AdS$_3$ solution and the BTZ black holes, TMG admits other vacuum known as warped AdS$_3$ \cite{Vuorio:1985ta,Percacci:1986ja} and its corresponding black holes \cite{Nutku:1993eb,Gurses,Bouchareb:2007yx}.

  Warped AdS$_3$ solutions can be obtained by fibring the real line over $AdS_2$ but with a constant warp factor multiplying the fiber metric. As a result, the $SL(2,R)\times SL(2,R)$ isometry of the AdS$_3$ breaks down to $SL(2,R)\times U(1)$. It was shown in \cite{Anninos:2008fx} that the warped AdS$_3$ black holes are  discrete  quotients of  warped AdS$_3$ by an element of $SL(2,R)\times U(1)$. Motivated by this fact it was conjectured in \cite{Anninos:2008fx} that the warped AdS$_3$ gravity has dual conformal field theory with  central charges
 \begin{equation}\label{central charges}
    c_L={4\nu\ell\over G(\nu^2+3)},\qquad\qquad c_R={\ell(5\nu^2+3)\over G\nu(\nu^2+3)}
 \end{equation}

Entropy calculation using Cardy formula and asymptotic symmetry group (ASG)  analysis \cite{Compere:2008cv}-\cite{Chen:2010qm} support this conjecture . Nevertheless,  the warped AdS$_3$ geometries do not have the full  $SL(2,R)\times SL(2,R)$ conformal  symmetry  in the bulk and this is somehow different from  the usual  AdS/CFT  dictionary. Recently, in an interesting paper \cite{Castro:2010fd} a similar problem has been studied in another geometry, namely  the 4D Kerr black hole.   According to \cite{Castro:2010fd} (which is  the continuation of \cite{Guica:2008mu} - \cite{ Matsuo:2009sj} ), for a AdS/CFT-type duality to work it is not necessary that the conformal symmetry has a realization in terms of the spacetime symmetries. Rather it is sufficient that
 the solution space of the wave equation for the propagating field has a conformal symmetry. Generalization of this idea to other black holes in 4D and 5D and the study of its  various  aspects  have been recently carried out in\cite{Krishnan:2010pv}-\cite{Wang:2010ic}.

In this paper we show that  a similar idea is at work  for  warped  $AdS_3$ black holes.  To do so,     similarly  to  \cite{Castro:2010fd}, we show that a massive  scalar field propagating in the spacelike warped AdS$_3$ black hole can probe a  hidden conformal symmetry. This hidden symmetry is unfolded  by reducing the scalar Laplacian to the Casimir of the $SL(2,R)$ symmetry. It is notable that this reduction is done for a particular limit of the scalar field frequency.
We find two sets of vector fields which identify the  $SL(2,R)\times SL(2,R)$  symmetry of the solution space. We should note that this symmetry is only locally defined and is broken down to  $U(1)\times U(1)$  by the  periodic identification of the $\phi$ coordinate. This is somehow similar to  AdS$_3$ case  which yields $BTZ$ black hole upon a discrete identification. Hence,  using the generator of the identification, we can read the left and right temperatures of the dual CFT which are in agreement with $WAdS/CFT$ conjecture \cite{Anninos:2008fx}.  As a result, massive  scalar fields with  frequencies less than a particular value do not see the  warped aspect of the background.  Moreover, the range of frequencies  which makes   the above reduction of the scalar wave equation possible, brings a significant simplification on the  absorption cross section of the scalar field and one can see remarkable  similarity between the final result and the two-point function of the  conjectured dual conformal field theory.

After a quick  review of  the warped AdS$_3$ black hole in section 2, in  section 3 we identify the hidden $SL(2,R)\times SL(2,R)$ symmetry and calculate the  absorption cross section of the scalar field and rewrite  it in terms of the CFT quantities. The last section is devoted to conclusions.

\section{Warped $AdS_3$ Black Holes }
 Topologically Massive Gravity (TMG) is given by the action
 \begin{equation}\label{TMGL}
 I_{TMG} = \frac{1}{16 \pi G} \left[\int_M  d^3x \, \sqrt{-g} (R+ \frac 2 {\ell^2}) + \frac{1}{\mu} \; I_{CS} \right],
\end{equation}
where $\mu$ is a positive constant and
\begin{equation}
 I_{CS} = \frac{1}{2} \int_M d^3x \, \sqrt{-g} \varepsilon^{\lambda \mu \nu} \Gamma^\alpha_{\lambda \sigma} \left(\partial_\mu \Gamma^\sigma_{\alpha \nu} + \frac{2}{3} \Gamma^{\sigma}_{\mu \tau}\Gamma^{\tau}_{\nu \alpha}\right)
\end{equation}
is the gravitational  Chern-Simons term \cite{Deser:1981wh,Deser:1982vy}.
The theory admits regular black hole solution for  \cite{Nutku:1993eb,Gurses,Bouchareb:2007yx}
\begin{equation}
   \nu\equiv{\mu\ell\over 3}>1.
\end{equation}
 These black holes may be  obtained by performing discrete identifications in the spacelike $WAdS_3$ given as \cite{Anninos:2008fx}
 \begin{equation}\label{warped ads metric}
    ds^2={\ell^2\over \nu^2+3}\left[-\cosh^2\sigma\,d\tau^2+d\sigma^2+{4\nu^2\over{\nu^2+3}}(du+\sinh\sigma d\tau)^2\right]
\end{equation}

In the ADM parametrization,  the spacelike warped $AdS_3$ black holes is given by
\be\label{warped BH metric} ds^2=-N(r)^2
dt^2+\ell^2R({r})^2(d\phi+N^\phi(r) dt)^2+{\ell^4d{r}^2\over4
R({r})^2N(r)^2}, \ee where
\begin{eqnarray}\label{def of metric functions}
\nonumber R({r})^2&\equiv&{{r}\over4}\left(3(\nu^2-1){r}+(\nu^2+3)({r}_++{r}_-)-4\nu\sqrt{{r}_+{r}_-(\nu^2+3)}\right),\\
\nonumber N({r})^2&\equiv&{\ell^2(\nu^2+3)({r}-{r}_+)({r}-{r}_-)\over4 R({r})^2},\\
N^\phi({r})&\equiv&{2\nu{r}-\sqrt{{r}_+{r}_-(\nu^2+3)}\over2R({r})^2}.
\end{eqnarray}
and  $r_+$ and $r_-$ are respectively the location of the outer and inner horizons.

For $\nu=1$ this metric reduces to the metric of BTZ in a rotating frame. Moreover, for $\nu<1$ the metric \eqref{warped BH metric} results in closed  timelike curves and hence we only consider the  $\nu\geq1$ range.

The Hawking temperature ,$T_H$, and the angular velocity at the outer horizon ,$\Omega_H$, are given by
\begin{eqnarray}\label{hawking and angular}
    T_{H}  &=& \frac{(\nu^2+3)(r_{+}-r_{-})}{4\pi\ell (2\nu r_{+}-\sqrt{(\nu^2+3)r_{+}r_{-}})},\\
    \Omega_{H}  &=& -\frac{2}{\ell(2\nu r_{+}-\sqrt{(\nu^2+3)r_{+}r_{-}})}.
\end{eqnarray}

One can easily verify that black holes \eqref{warped BH metric} satisfy the first law of thermodynamics for the values of the Wald entropy, ADT mass and angular momentum given by \cite{Abbott:1982jh}-\cite{Deser:2003vh}
\begin{equation}
S={\pi\ell\over24\nu G}\left[(9\nu^2+3){r}_+-(\nu^2+3){r}_--4\nu\sqrt{(\nu^2+3){r}_+{r}_-}\right],
\end{equation}
\begin{equation}
\mathcal{M}^{ADT} =  {(\nu^2+3)\over24 G}\left(\nuo\right),
\end{equation}
\begin{multline}
\mathcal{J}^{ADT} =  \frac{\nu\ell(\nu^2 + 3)}{96 G}\left[\left(\nuo\right)^2\right.
\left.- \frac{(5\nu^2+3)}{4\nu^2}({r}_+-{r}_-)^2\right].
\end{multline}

\section{Hidden conformal   Symmetry}
In this section we consider a massive  scalar field propagating in the background of  warped AdS$_3$ black hole \eqref{warped BH metric}. The classical wave equation describing  the dynamics of the  scalar  field is
  \be\label{KG
equation} \left({1\over
\sqrt{-g}}\,\partial_\mu\sqrt{-g}\partial^\mu-m^2\right)\Phi=0.
\ee By making use of the Fourier expansion
\begin{equation}
    \Phi(t,r,\phi)=e^{-i\omega t+i{k}\phi}S(r)
\end{equation}
and  using \eqref{warped BH metric}-\eqref{def of metric functions}, the
equation \eqref{KG equation}  can be recast into
\begin{equation}\label{scalar eq before limit}
\begin{split}
    &\partial_r\left((r-r_+)(r-r_-)\partial_r
    S(r)\right)+
    {\left(2\nu r_+ \omega-\omega\sqrt{r_+r_-(\nu^2+3)}
    +2k\right)^2\over(r-r_+)(r_+-r_-)(\nu^2+3)^2}S(r)\\&-{\left(2\nu r_- \omega-\omega\sqrt{r_+r_-(\nu^2+3)}
    +2k\right)^2\over(r-r_-)(r_+-r_-)(\nu^2+3)^2}S(r)+{3(\nu^2-1)\over(\nu^2+3)^2}\omega^2S(r)={\ell^2m^2\over \nu^2+3} S(r)
\end{split}
\end{equation}
 Solutions to this equation have been previously  studied  in literature  (see \cite{Oh:2008tc}-\cite{Chen:2009cg}). In this  section we will observe  that for
the cases that the last term of \eqref{scalar eq before limit} can
be neglected, the scalar Laplacian of the wave equation can be
written as an $SL(2,R)$ Casimir. For $\nu=1$ the last $\omega^2$ term of
\eqref{scalar eq before limit} is zero and no extra condition is
needed. Since for $\nu=1$ the
warped black holes  reduce to the BTZ black holes in the
rotating coordinate, this is consistent with the local $SL(2,R)\times SL(2,R)$ symmetry of the BTZ background.

In order to neglect the last term of  \eqref{scalar eq before limit}
for $\nu>1$, we require that\footnote{It is clear from \eqref{warped
BH metric} that $t$ is dimensionless and hence $\omega$ is
dimensionless too. The natural units for energy is $1/l$. }
\begin{equation}\label{condition on omega}
    \omega^2\ll{(\nu^2+3)^2\over 3(\nu^2-1)}.
    \end{equation}
Unlike the higher dimensional black holes \cite{Castro:2010fd},\cite{Krishnan:2010pv}-\cite{Chen:2010zw}, it is not
necessary to divide the geometry to the near and the far regions. As
we will see   , condition \eqref{condition on
omega} is enough to reduce the scalar Laplacian of equation
\eqref{scalar eq before limit} to the $SL(2,R)$ Casimir. This means
that a scalar field with sufficiently small   frequency  can probe the $SL(2,R)$
symmetry everywhere in the spacetime.

As has been done in \cite{Anninos:2009jt} and \cite{Chen:2009hg}, one may propose that instead of imposing  \eqref{condition on omega}, it is possible to include  the last $\omega^2$ term to the right hand side of \eqref{scalar eq before limit} and write the remaining part of the  LHS of \eqref{scalar eq before limit}  as the casimir of $SL(2,R)$. However, then  because of the new $\omega^2$  term in the RHS, the highest conformal weights will be  frequency-dependent. We need to  avoid this frequency-dependent  weights in the context of  hidden symmetry  because, as we will see explicitly in the next subsection, $\omega$  is the eigenvalue of a non-Casimir combination of $SL(2,R)$ generators. Hence, the $SL(2,R)\times SL(2,R)$ symmetry of \eqref{scalar eq before limit} is only manifest in the low energy limit set by \eqref{scalar eq before limit}.

Let us try to understand  the condition  \eqref{condition on omega} better. Noting that the Warped AdS$_3$ space \eqref{warped ads metric} and Warped AdS$_3$ black hole \eqref{warped BH metric} are related by a local coordinate transformation \cite{Anninos:2008fx} (see \eqref{local coord trans} ), it will prove convenient to translate \eqref{condition on omega} in terms of the frequency and wavelength of the scalar field propagating in the Warped AdS$_3$ space \eqref{warped ads metric}. One simple way to do so is using the identification of \cite{Chen:2009hg} to relate the $(\omega,k)$ of the warped black hole to the $(\tilde\omega,\tilde k)$ of the warped AdS$_3$ background:
\begin{equation}\label{relation between parameters}
    \tilde\omega={2\over \nu^2+3}k,\qquad \tilde k={2\nu\over\nu^2+3}\omega
\end{equation}
In fact, this relations has been introduced in \cite{Chen:2009hg} by using the identification of \cite{Anninos:2009zi} to relate the asymptotic geometry of the warped AdS$_3$ space and the warped AdS$_3$ black hole.

In terms of the parameters of the warped AdS$_3$ background, condition \eqref{condition on omega} takes the  form
\begin{equation}\label{condition on warped ads}
    {1\over\tilde k^2}\gg{3(\nu^2-1)\over 4\nu^2}=1-\alpha,
\end{equation}
where $\alpha$ is the  ratio of the radius of the AdS$_2$ space to  the radius of the  fiber space in \eqref{warped ads metric}. For the ordinary AdS$_3$  space, $\alpha=1$  and the deviation from this value indicates the amount of warping.    Condition \eqref{condition on warped ads} shows that in order to have hidden $SL(2,R)$ symmetry, the wavelength of the scalar field must be much greater than the difference between the radius of the AdS$_2$  and the radius of the  fiber space in \eqref{warped ads metric}, so that the warping essentially remains unnoted by the scalar probe\footnote{The author would like to thank M. M. Sheikh Jabbari for making this point.}.

\subsection{$SL(2,R)_L\times SL(2,R)_R$ symmetry}

Let us now define the vector fields
\begin{equation}\label{first sl2R}
 \nonumber H_0 = -{2i\nu\over \nu^2+3}{T_L\over T_R}\partial t+{i\over 2\pi\ell  T_R}\partial_\phi,
 \end{equation}
 \begin{equation}
 \begin{split}
 \nonumber H_1 = i\,e^{-2\pi\ell T_R\phi}\Bigg[-{\nu\over(\nu^2+3) \sqrt{\Delta}}&\left((2r-r_+-r_-){T_L\over T_R}+r_+-r_-\right)\partial_t+\sqrt{\Delta}\partial_r\\
   &+{2r-r_+-r_-
 \over 4\pi\ell T_R\sqrt{\Delta}}\partial_\phi\Bigg ],
 \end{split}
 \end{equation}
\begin{equation}
 \begin{split}
  H_{-1} = i\,e^{2\pi\ell T_R\phi}\Bigg[-{\nu\over(\nu^2+3) \sqrt{\Delta}}&\left((2r-r_+-r_-){T_L\over T_R}+r_+-r_-\right)\partial_t-\sqrt{\Delta}\partial_r\\
   &+{2r-r_+-r_-
 \over 4\pi\ell T_R\sqrt{\Delta}}\partial_\phi\Bigg ],
 \end{split}
 \end{equation}
and also
\begin{eqnarray}\label{second sl2R}
 \nonumber \bar H_0 = {2i\nu\over \nu^2+3}\partial t,
 \end{eqnarray}
 \begin{equation}
 \begin{split}
 \nonumber \bar H_1 = i\,e^{-{\nu^2+3\over 2\nu}t-2\pi\ell T_L\phi}\Bigg[{\nu\big((\nu^2+3)(2r-r_+-r_-)+8\pi\ell T_L\big)\over(\nu^2+3)^2\sqrt{\Delta}}\partial_t
 &+\sqrt{\Delta}\partial_r\\&-{2\over(\nu^2+3)\sqrt{\Delta}}\partial_\phi \Bigg],
 \end{split}
 \end{equation}
 \begin{equation}
 \begin{split}
  \bar H_{-1} = i\,e^{{\nu^2+3\over 2\nu}t+2\pi\ell T_L\phi}\Bigg[{\nu\big((\nu^2+3)(2r-r_+-r_-)+8\pi\ell T_L\big)\over(\nu^2+3)^2\sqrt{\Delta}}\partial_t
 &-\sqrt{\Delta}\partial_r\\&-{2\over(\nu^2+3)\sqrt{\Delta}}\partial_\phi \Bigg],
 \end{split}
 \end{equation}
where
\begin{equation}
     T_R = {(\nu^2+3)(r_+-r_-)\over 8\pi\ell},
 \end{equation}
 \begin{equation}
    T_L ={\nu^2+3\over 8\pi\ell}\left(r_++r_--{\sqrt{r_+r_-(\nu^2+3)}\over \nu}\right),
  \end{equation}
  \begin{equation}
        \Delta = (r-r_+)(r-r_-).
\end{equation}
It is not difficult to see that they satisfy the $SL(2,R)\times SL(2,R)$ algebra
\begin{eqnarray}
 \nonumber [H_n,H_m] &=& i(n-m)H_{n+m} \\
 \nonumber {[}\bar {H}_n,\bar{H}_m{]}   &=& i(n-m)\bar {H}_{n+m} \quad\qquad\qquad (n,m=-1,0,1)\\
  {[}H_n,\bar H_m{]} &=&0
\end{eqnarray}
Moreover, one may  readily  verify that after imposing condition \eqref{condition on omega} the wave equation \eqref{scalar eq before limit}  can be written as
\begin{equation}\label{casimir equation}
    \mathcal{H}^2\Phi= \mathcal{\bar H}^2\Phi={\ell^2m^2\over \nu^2+3}\Phi
\end{equation}
where $\mathcal{H}^2$ and $\mathcal{\bar H}^2$ are the $SL(2,R)$ quadratic Casimir given by
\begin{equation}\label{def  casimir}
    \mathcal{H}^2=-H_0^2+{1\over2}(H_1H_{-1}+H_{-1}H_1),
\end{equation}
and similarly for $\mathcal{\bar H}^2$. Hence the scalar Laplacian can be written as the $SL(2,R)$ Casimir and  the $SL(2,R)_L\times SL(2,R)_R$ weights of the scalar field are
\begin{equation}\label{weights}
    (h_L,h_R)=(\frac12\sqrt{1+{4\ell^2m^2\over \nu^2+3}}-\frac12,\frac12\sqrt{1+{4\ell^2m^2\over \nu^2+3}}-\frac12).
\end{equation}
It is notable that this conformal weights are in agreement  with the results of \cite{Anninos:2009jt} and \cite{Chen:2009hg} in the low frequency limit  \eqref{condition on omega}.

Similar to higher dimensional black holes \cite{Castro:2010fd},\cite{Krishnan:2010pv}-\cite{Chen:2010zw}, the above
$SL(2,R)\times SL(2,R)$  symmetry  acts  on the solution
space only locally because  the vector fields \eqref{first sl2R} and
\eqref{second sl2R} are not periodic under the identification
$\phi=\phi+2\pi$. This identification is generated by the group
element
\begin{equation}\label{identification}
    e^{\partial_\phi }=e^{-i2\pi\ell(T_RH_0+T_L\bar H_0)}
\end{equation}
which is the same identification used in \cite{Anninos:2008fx} to produce the warped
black holes as a quotient of the warped $AdS_3$. Interestingly,
$H_0$ and $\bar H_0$ coincide respectively  with  $-i{\tilde J_2/
2}$ and $i{ J_2/ 2}$ of \cite{Anninos:2008fx} and $T_R$ and $T_L$ with the
advertised right and left temperatures of the  conformal field
theory dual to the warped $AdS$ black holes. Thus  we obtain
another evidence for  the conjecture of \cite{Anninos:2008fx} that  warped
$AdS_3$ gravity of TMG has a holographic dual description in terms
of a 2 dimensional conformal field theory. The Entropy calculation
of \cite{Anninos:2008fx} using the Cardy formula
\begin{equation}\label{cardy }
    S={\pi^2\ell\over 3}(c_LT_L+c_RT_R)
\end{equation}
supports this hypothesis.

Now let us  define a local coordinate transformation as
\begin{eqnarray}\label{local coord trans}
  \nonumber \tau &=& \tan^{-1}\left({2\sqrt\Delta\over 2r-r_+-r_-}\sinh\left(2\pi \ell T_R\phi\right)\right),\\
  \nonumber u &=& {\nu^2+3\over 2\nu} t+2\pi\ell T_L\phi+\tanh ^{-1}\left({2r-r_+-r_-\over r_+-r_-}\coth(2\pi\ell T_R\phi)\right),\\
  \sigma &=& \sinh^{-1}\left({2\sqrt\Delta\over r_+-r_-}\cosh(2\pi\ell T_R\phi)\right).
 \end{eqnarray}
In this new coordinate the metric of warped AdS$_3$ black hole \eqref{warped BH metric} takes the form \eqref{warped ads metric}  i.e. the metric of the spacelike warped AdS$_3$ geometry \cite{Anninos:2008fx}. In the coordinate $\{\tau,\sigma,u\}$, the vector fields $H_n$ and $\tilde H_n \,(n=-1,0,1)$ are written  as
\begin{equation}\label{H in new coord}
    H_0=-{i\over2}{\tilde J}_2,\qquad H_1={i\over2}(\tilde {J}_0-\tilde {J}_1),\qquad H_{-1}={i\over2}(\tilde J_0+\tilde J_1),
\end{equation}
\begin{equation}\label{tild H in new coord}
    {\tilde H}_0={i\over2}{ J}_2,\qquad {\tilde H}_1={i\over2}( {J}_1- {J}_0),\qquad {\tilde H}_{-1}={i\over2}( J_1+ J_0),
\end{equation}
where
\begin{eqnarray}
J_1 &=& {-\frac{2 \sinh{u}}{\cosh{\sigma}} \partial_{\tau}-2\cosh{u} \partial_{\sigma}+2\tanh{\sigma} \sinh{u} \partial_{u}}, \\
J_2 &=& 2\partial_{u}, \\
J_0 &=& {\frac{2 \cosh{u}}{\cosh{\sigma}} \partial_{\tau}+2\sinh{u} \partial_{\sigma}-2\tanh{\sigma} \cosh{u} \partial_{u}},
\end{eqnarray}
and
\begin{eqnarray}
\tilde{J}_1 &=&  {2\sin{\tau} \tanh{\sigma} \partial_{\tau} - 2\cos{\tau} \partial_{\sigma}+\frac{2\sin{\tau}}{\cosh{\sigma}} \partial_{u} },  \\
\tilde{J}_2 &=&  {-2\cos{\tau} \tanh{\sigma} \partial_{\tau}-2\sin{\tau} \partial_{\sigma} - \frac{2\cos{\tau}}{\cosh{\sigma}} \partial_{u}  },  \\
\tilde{J}_0 &=&  2\partial_{\tau},
\end{eqnarray}
are the $SL(2,R)_L\times SL(2,R)_R$ Killing vectors of an AdS$_3$  written in the fibred form
\begin{equation}\label{ ads metric}
    ds^2={\ell^2\over 4}\left[-\cosh^2\sigma\,d\tau^2+d\sigma^2+(du+\sinh\sigma d\tau)^2\right].
\end{equation}
This is  interesting  that the hidden conformal symmetry of the spacelike warped AdS$_3$ black hole is locally the isometry of the  AdS$_3$ geometry. This means that scalar fields with frequency \eqref{condition on omega} do not feel   the warped  feature of the spacetime.

\subsection{Absorption cross section}
In order to  elaborate on the hidden aspect of the  conformal symmetry, we consider scattering of the massive  scalar field in the background of the  warped AdS$_3$ black hole.  It is notable that this issue has recently  been studied  in \cite{Oh:2008tc,Oh:2009if,Kao:2009fh}. In this section we only clarify the effect of the condition \eqref{condition on omega} in their calculations.

 After imposing condition \eqref{condition on omega} on the scalar wave equation \eqref{scalar eq before limit}, the absorption cross section of the scalar field   takes the following  form (see formula (5.1) of \cite{Kao:2009fh})

\begin{eqnarray}\label{abs}
 \nonumber \sigma_{abs}
  %&\sim&  |A|^{-2} \\
 \nonumber  &\sim&
   \sinh\bigg[2\pi\big(\Omega_+\omega+Uk\big)\bigg]\Bigg|\Gamma\bigg(\frac12+\frac12\sqrt{1+{4m^2\ell^2\over{\nu^2+3}}}-i\big(\omega(\Omega_++\Omega_-)+2Uk\big)\bigg)\Bigg|^2\\
   &&\qquad\times\Bigg|  \Gamma\bigg(\frac12+\frac12\sqrt{1+{4m^2\ell^2\over{\nu^2+3}}}-i\omega\big(\Omega_+-\Omega_-\big)\bigg) \Bigg|^{2}
\end{eqnarray}
where\begin{eqnarray}
  U &=& {2\over (r_+-r_-)(\nu^2+3)} \\
  \Omega_+ &=& {2\nu r_+-\sqrt{r_+r_-(\nu^2+3)}\over (r_+-r_-)(\nu^2+3)} \\
  \Omega_- &=& {2\nu r_--\sqrt{r_+r_-(\nu^2+3)}\over (r_+-r_-)(\nu^2+3)}
\end{eqnarray}

Now we want to rewrite \eqref{abs} in terms of the dual CFT parameters such as left and right temperatures, their conjugate charges and the conformal weights $(h_L,h_R)$. The conjugate charges $\delta E_L$ and $\delta E_R$ are determined through the relation
\begin{equation}\label{first law2}
    \delta S={\delta E_L\over T_L}+{\delta E_R\over T_R}
\end{equation}
where $S$ is the entropy. Using the first law of thermodynamics
\begin{equation}\label{fist law 1}
     \delta S={\delta M\over T_H}-{\Omega_H\over T_H}{\delta J}
\end{equation}
and identifying $\delta M=\omega/\ell$ and $\delta J= k$, we can find the left and right conjugate charges as
\begin{eqnarray}
  \omega_L &\equiv& \delta E_L= {1\over 2\ell}\left(\nu({r}_++{r}_-)-\sqrt{{r}_+{r}_-(\nu^2+3)}\right)\omega \\
  \omega_R &\equiv& \delta E_R= {1\over 2\ell} \left(\nu({r}_++{r}_-)-\sqrt{{r}_+{r}_-(\nu^2+3)}\right)\omega+{k\over\ell}
\end{eqnarray}
Finally  we can rewrite \eqref{abs} as

\begin{equation}\label{CFT result}
    {\sigma_{\rm abs}\sim T_L^{2h_L-1}T_R^{2h_R-1}\sinh\left({\omega_L\over 2T_L}+{\omega_R \over 2 T_R}\right)\left|\Gamma(h_L+i{\omega_L \over 2\pi T_L})\right|^2\left|\Gamma(h_R+i{\omega_R\over 2\pi T_R})\right|^2~}.
\end{equation}
which is the well-known absorption cross section for a 2d CFT.

\section{Conclusion}
In this letter we have  studied the hidden conformal symmetry of  warped AdS$_3$ black holes.  We have shown that although the  warped AdS$_3$ background does not have a $SL(2,R)_L\times SL(2,R)_R$ isometry, it is still the hidden symmetry of the solution space . The context is very similar to that in   higher dimensional black holes and seems to open a new window into    the BH/CFT duality.  Although  we have only considered  scattering of a scalar field, it is plausible that the analysis  works for any other fields  propagating in the  warped AdS$_3$ background. Our result may be thought  of as a supporting evidence in favor  of the WAdS/CFT conjecture.

 An important question at this stage  is  how this hidden symmetry  can be enhanced to the Virasoro algebra.  In fact this question is in the light of   Brown and Henneaux's work \cite{Brown:1986nw} that the local  symmetry of the background is enhanced to the Virasoro algebra at the boundary by making use of the {\textit{asymptotic symmetry group}}. Generalizing the $ASG$  method for the hidden symmetry of the bulk and using it to identify somehow the  central charges of the dual CFT seem interesting questions.

\section*{Acknowledgments}
The author would like to thank A. Alishahiha, M. M. Sheikh Jabbari and A. E. Mosaffa for  valuable discussions and  comments on the manuscript. We would also like to thank S. Detournay and M. Vincon for useful comments on revised version.

%--------------------------------------------------------------------

%---------------------------------------------------------------------
%Bibliography

%--------------------------------------------------------------------
\end{document}